\documentclass[useAMS,usenatbib,usegraphicx]{mn2e}

\title[A stability catalogue of the habitable zones]
{A stability catalogue of the habitable zones in extrasolar planetary systems}

\author[S\'andor et al.]
{Zs. S\'andor,$^1$ \'A. S\"uli,$^1$ B. \'Erdi,$^1$\thanks{Corresponding author: B.Erdi@astro.elte.hu}  E. Pilat-Lohinger$^2$ and R. Dvorak$^2$ \\
$^1$Department of Astronomy, E\"otv\"os University, P\'azm\'any P\'eter s\'et\'any 1/A, H-1117 Budapest, Hungary\\
$^2$Institute for Astronomy, University of Vienna, T\"urkenschanzstrasse 17, A-1180 Vienna, Austria}

\date{Released 2002 Xxxxx XX}

\begin{document}

\label{firstpage}

\maketitle
\begin{abstract}
In the near future there will be launched space missions (e.g. COROT, KEPLER), designed to detect Earth-like extrasolar planets. The orbital elements of these (still hypothetic) planets will contain some uncertainties, that can only be eliminated by careful  dynamical investigations of the hosting planetary systems. The proportion of extrasolar planetary systems with one known giant planet is high ($\sim 90 \%$), therefore as a first step we have investigated the possible existence of terrestrial planets in these systems.  In this paper a development of a stability catalogue of the habitable zones of exoplanetary systems is reported. This catalogue is formed by a series of stability maps, which can help to establish, where Earth-like planets could exist in extrasolar planetary systems having one giant planet. After a description of the dynamical model and the numerical methods, details of the stability maps are discussed. An application of the stability catalogue to 15 known exoplanetary systems is also shown, and a characterization of the stability properties of their habitable zones is given.
\end{abstract}

\section{Introduction}

Since the discovery of the first extrasolar planet around a solar-type star, 51 Pegasi by \citet{may95}, by now (October 2006) 209 exoplanets have been found. These planets form 179 exoplanetary systems, among which there are 21 multiple systems. The main detection technique of exoplanets is based on the Doppler shift of the hosting star's spectral lines. This method is adequate to discover giant planets, but it is not yet capable to detect Earth-sized planets.

However, the detection and investigation of Earth-like exoplanets can be of great importance for the further improvements of formation theories of our Solar System, where both rocky Earth-like, and gaseous Jupiter-like planets have been formed. Another major question which has a special interest not only for the scientific community is whether life may have evolved in some exoplanetary systems. The region of an exoplanetary system, where water-based life might evolve, is called the habitable zone (HZ). Simply saying, the HZ of a planetary system is that region, where liquid water can exist on the surface of a planet \citep{kas93}. The dynamical stability of hypothetical Earth-like planets in the HZs of several extrasolar planetary systems has been the subject of intensive research \citep*[see for instance][]{jon01,men03,asg04,jon05,jon06}. According to these studies many systems can host terrestrial exoplanets in their HZs.

Recently, there have been much efforts for the construction of space probes aiming at the detection of terrestrial exoplanets in the near future. Such is COROT, to be launched in November 2006, and KEPLER with a launch in 2008. These missions will use transit photometry as detection technique. The transit of an unseen exoplanet in front of the hosting star's disc is expected to result in an observable dimming of the star's light intensity. There are also other plans, such as NASA's TPF and SIM, or ESA's Darwin and Gaia missions in the next decade aiming at studying all aspects of exoplanets.

By using data obtained from the observations, the orbital parameters of an exoplanet can be derived only within some error limits. In order to improve the orbital parameters certain constraints should be used. For instance, if an Earth-like planet were discovered in an extrasolar planetary system, where an already detected giant planet is known to orbit the star, such a constraint could be the dynamical stability of the terrestrial planet. This means, that those orbital solutions for the terrestrial planet, which result in unstable orbits, are unlikely and therefore should be excluded.

The dynamical stability of Earth-like planets can be established in several ways, for example (i) by exploring the stable and unstable regions of the phase space of each extrasolar planetary system separately, or (ii) by using stability maps computed in advance for a large set of orbital parameters. In this paper we present such stability maps and show how to apply them to the detected exoplanetary systems. The second method has the advantage that the stability properties of a terrestrial planet can be easily established when the orbital parameters of the giant planet of an exoplanetary system are modified. Very often the orbital parameters of the giant planets are uncertain, and due to the accumulation and improvements of the observational data, they are changed quite frequently. If the first approach is chosen, one has to re-explore the phase space of the individual exoplanetary system after each modification of the orbital parameters of the giant planet. However, this is not necessary when the second method is used, since the stability properties of the investigated planetary system can be easily re-established from the already existing stability maps.

As a first step towards general stability maps, we compiled a stability catalogue for a dynamical model consisting of a star, a giant planet, and a small Earth-like planet. We present this catalogue in this paper. Section 2 describes the dynamical model and the applied numerical methods. Details of the catalogue and the stability maps are given in Sections 3 and 4. Section 5 shows how to apply the catalogue to individual exoplanetary systems.
By using this catalogue, the dynamical reality of the orbital elements of a newly discovered terrestrial planet in an extrasolar planetary system can be easily determined. This catalogue can also be used to establish the stability properties of the HZs of known exoplanetary systems. This is shown in Section 6.

\section{Dynamical model and methods of investigation}

The majority of planetary systems, detected up to now, consists of a star and a giant planet revolving about it in an eccentric orbit. Therefore we use a relatively simple dynamical model (known as the elliptic restricted three-body problem), where two massive bodies, called the primaries, move in elliptic orbits about their common center of mass, and a third body of negligible mass moves under their gravitational influence  \cite[for details see][]{szeb67}. In our application the primaries are a star and a giant planet, and the third body is a small Earth-like planet, regarded as massless.

We note, however, that among the extrasolar planetary systems with long observational baselines, there is a high rate of multiple planet systems. Thus a more convenient model for our stability investigations would be the restricted $N$-body problem with $N-2$ giant planets ($N \geq 4$). The presence of additional giant planets may enhance the instabilities, induced by just one massive planet, and make the HZ of a system more unstable. The instabilities may appear due to mean motion resonances between the massive giant planet and the Earth-like planet. Thus, by mapping these resonances, we can find the most significant birthplaces of instabilities in the habitable zones. On the other hand, the dynamical model with one giant planet also offers the most convenient way to display the most important mean motion resonances as a function of the mass ratio of the star and the giant planet, and of the eccentricity of the giant planet. In this paper we mapped these resonances, and the effects of additional giant planets will be studied in a future work.

To explore the stable and unstable regions of the motion of the third body, the equations of motion have to be integrated for a large set of initial conditions and for a long time for each orbit. This can be very time consuming, even in the case of a huge computing potential. In order to study efficiently the stability properties of dynamical systems, several numerical techniques and chaos detection methods have been developed recently. Using the working hypothesis, that chaotic behaviour means instability through chaotic diffusion of trajectories in the phase space, these methods can be applied to explore the stable and unstable regions of dynamical systems. We are, however, aware of the fact, that chaotic behaviour does not necessarily mean instability. There could be orbits exhibiting chaotic behaviour, but stable in Nekhoroshev-sense, e.g. remaining in a bounded region of the phase space for time intervals equal to the age of the investigated planetary system. However, if other destabilizing effects, such as the presence of yet undetected giant planets are present in the exoplanetary systems, chaotic behaviour may result in really unstable orbits. In other words, in high dimensional dynamical systems unstable orbits emanate only from the chaotic regions of the phase space. Exploring these regions we can find the birth domains of unstable orbits. However, in order to decide the stability character of an orbit, one has to use long-term numerical integration of the equations of motion. On the other hand, in the case of a large set of initial conditions the fast chaos detection methods allow us to explore qualitatively the dynamical structure of the extrasolar planetary systems.

Our investigations are based on such chaos detection methods. As a main tool we used the method of the relative Lyapunov indicators (RLI) \citep{san00, san04}, and in some cases the fast Lyapunov indicators (FLI) \citep{fro97, guz02}. We also applied the maximum eccentricity method (MEM) \citep[see for instance][]{dvor03}, as an independent check of the RLI and the FLI by direct numerical integration of the equations of motion. The knowledge of the eccentricity of stable orbits is essential for the problem of the development of a biosphere on a terrestrial planet in the habitable zone of a star. For example, recent estimations for the maximum eccentricity for the Earth -- after detailed atmosphere models\footnote{Yu. N. Kulikov (personal communication 2nd ISSI team meeting, Bern, 20-22.2.2006: Evolution of Habitable Planets)} -- are $e_{max}<0.2$. 

Regarding the RLI and the FLI, our choice is motivated by the complementary character of the two chaos detection methods. The RLI measures the difference between the convergence of the finite time Lyapunov indicators to the maximal Lyapunov characteristic exponent of two initially very close orbits. This method is extremely fast in determining the ordered or chaotic nature of individual orbits, even to detect orbits with very small Lyapunov characteristic exponents  \cite[e.g. sticky orbits,][]{san04}, as well as to distinguish between ordered and chaotic regions of the phase space. According to our experiments, gained in different dynamical problems, it is enough to integrate the two initially close orbits for a few hundred times of the longest orbital period of the investigated system. As to the FLI, their time evolution shows clearly the ordered or chaotic character of an orbit, and the method is very effective in detecting stable resonant orbits. Concerning the computation of the FLI, the integration time is several tens of thousands orbital periods. In the MEM we compute by numerical integration the largest value of the orbital eccentricity, reached by an orbit during a given, long time interval ($10^5$ orbital periods in this paper). These methods have been successfully applied to several exoplanetary systems  to determine their dynamical stability properties: for the RLI see \citep{erd04,san04}, for the FLI  \citep{plo02,dvor03,boi03}, and for the MEM \citep{dvor03,erd04}.

\section{Catalogue of dynamical stability}

We compiled a catalogue of dynamical stability, adapted to a system consisting of a star with mass $m_0$, a giant planet with mass $m_1$, and a small test planet with negligible mass. The total mass of the star and the giant planet is taken as unit mass, thus $\mu=m_1/(m_0+m_1)$ is the mass parameter of the problem, which has been changed between (i) $1\times 10^{-4}$ and $9 \times 10^{-4}$ in steps of $\Delta\mu=10^{-4}$, (ii) between $1\times 10^{-3}$ and $9 \times 10^{-3}$ in steps of $\Delta\mu=10^{-3}$, and (iii) between $1\times 10^{-2}$ and $5 \times 10^{-2}$ in steps of $\Delta\mu=10^{-2}$. This results in 23 values for $\mu$. The giant planet was assumed to move around the star in an elliptic orbit, whose plane is taken as reference plane, with semi-major axis $a_1$, eccentricity $e_1$, argument of periastron $\omega_1$, and mean anomaly $M_1$. Without the loss of generality, the semi-major axis $a_1$ can be taken as unit distance, thus $a_1=1$. We assumed that $\omega_1=0^{\circ}$. The eccentricity $e_1$ was changed between 0.0 and 0.5 with a stepsize of $5 \times 10^{-3}$. During our computations we also changed $M_1$ between $0^\circ$ and $360^\circ$ with $\Delta M_1=45^\circ$ both for inner and outer orbits of the test planet (with respect to the orbit of the giant planet). Additionally, we changed $M_1$ for inner orbits of the test planet between $0^\circ$ and $90^\circ$ with $\Delta M_1=10^\circ$ too. We assumed that the test planet moves in the orbital plane of the giant planet, and we always took $e=0$, $\omega=0^{\circ}$, $M=0^{\circ}$ as initial orbital elements of the test planet. The case when the orbit of the test planet is inclined with respect to the giant planet's orbit will also be studied in future work. We changed the semi-major axis $a$ of the test planet in two different intervals: (a) for orbits of the test planet ``inside" the orbit of the giant planet we changed $a$ between 0.1 and 0.9 with a stepsize of $10^{-3}$, and (b) for ``outside" orbits between 1.1 and 4.0 with a stepsize of $3.625 \times 10^{-3}$.

The catalogue consists of 552 stability maps, corresponding to each ($\mu,M_1$) pair of the 23 values of $\mu$ and 16 values of $M_1$ for inner orbits (368 maps) and 8 values of $M_1$ for outer orbits (184 maps) of the test planet. For a given ($\mu,M_1$) pair the stability map is prepared in such a way, that for each set of the described initial conditions the RLI (or FLI) value of the corresponding orbit of the test planet is computed and visualized in the ($a,e_1$) parameter plane. Due to the very small stepsize in $a$ and $e_1$, each stability map corresponds to more than $8 \times 10^{4}$ initial conditions, thus providing a very fine resolution. In the following section we show some stability maps from the catalogue, containing the HZs of some exoplanetary systems as examples. The complete catalogue is available at \texttt{http://astro.elte.hu/exocatalogue} and \texttt{http://www.astro.univie.ac.at/adg}.

\section{Details of the stability maps}

In Fig. \ref{fig1} we displayed a case, where the semi-major axis of the test planet's orbit is larger than that of the giant planet ($a>1$, and $\mu=0.001$, $M_1=0^{\circ}$). Light regions correspond to low values of the RLI ($10^{-10}$), and thus to ordered, dynamically stable motion of the test planet. Dark shades correspond to large values of the RLI ($10^{-5}$) and chaotic behaviour of the test planet. The dark unstable regions on the left are due to the proximity of the giant planet. The figure is dominated by V-shaped gray stripes, corresponding to outer mean motion resonances between the test planet and the giant planet. These resonances, marked at the top of the figure, can represent either ordered (stable for infinite time), or weakly chaotic (which may become unstable after very long time) behaviour. With the increase of $e_1$ many resonances overlap, giving rise to strongly chaotic and thus very unstable behaviour. The reason for this is that by increasing $e_1$ the apocenter distance of the giant planet also increases, and the giant planet perturbs more strongly the outer test planet. The figure also shows, for two planetary systems, the positions of the zero age HZ, which we call the \emph{classical} HZ, and the \emph{present} HZ. We recall, that the habitable zone (HZ) is that region around a star, where water can exist in liquid phase on the surface of a planet by virtue of stellar irradiance. The boundaries of the classical HZ are calculated by assuming the physical properties of a zero age main sequence star, while the boundaries of the present HZ depend also on the age of the hosting star \citep{jon06}. The borders of the  classical HZs for a large number of extrasolar planetary systems are given by \cite{men03}, while the borders of the present HZs can be found in the more recent work of \cite{jon06}. We note that according to the latter work the present HZs are shifted outwards with respect to the location of the classical HZs.

Fig. \ref{fig2} shows a stability map for the same initial conditions as in Fig. \ref{fig1}, but resulting from the FLI computations. It has the same overall structure, as the map obtained by the RLI. Note, however, the complementary character of the results obtained by the two methods. The RLI detects the chaotic separatrices of the resonances (dark V-shaped stripes in Fig. \ref{fig1}), while the FLI finds the stable resonant orbits inside the resonances (light stripes inside the resonances in Fig. \ref{fig2}). Note, that for the given initial conditions ($M_1=0^{\circ}$) almost all resonances can have stable orbits, but the 1:3 resonance is chaotic according to both methods, a well known result for the 3:1 inner resonance \citep{wis83}.

Fig. \ref{fig3} has been calculated by the MEM, again for the same initial conditions as in Figs. \ref{fig1} and \ref{fig2}. It can be seen, that similarly to the RLI and the FLI, the MEM also indicates the mean motion resonances. Since in the MEM orbits were integrated for $10^5$ periods of the giant planet, we used it as a test of the two other chaos detection methods.

Figs. \ref{fig4} and \ref{fig5} show stability maps for outer orbits of the test planet, computed by the RLI for $\mu=0.002$ and $\mu=0.007$, respectively. It can be seen, that for larger values of $\mu$ the chaotic regions are more extended, due to the stronger perturbing effect of the giant planet.

In Figs. \ref{fig6} --\ref{fig10} we display cases where the semi-major axis of the test planet is smaller than that of the giant planet ($a<1$).  Two panels for $\mu=0.001$ (Figs. \ref{fig6}a and \ref{fig6}b) show the results for two different starting positions of the giant planet, $M_1=0^{\circ}$ (Figs. \ref{fig6}a) and $M_1=180^{\circ}$ (Figs. \ref{fig6}b ). Here again mean motion resonances, this time inner ones, between the test planet and the giant planet dominate the stability maps. Inside the resonances the stable or chaotic behaviour of the test planet depends on the initial angular positions of the two planets. This is clearly visible, when we compare the two panels. In Fig. \ref{fig6}a the giant planet starts from the pericenter, in Fig. \ref{fig6}b  from the apocenter, while the initial position of the test planet in both cases is at $M=0^\circ$, corresponding to the direction of the giant planet's pericenter. Comparing the two figures it can be seen that the location of the resonances does not change, since this depends on the ratio of the semi-major axes of the two planets. However, the character of the resonances becomes different.  In Fig. \ref{fig6}b several resonances (5:2, 5:3, 3:2) are more stable than in Fig. \ref{fig6}a. This is due to the fact that the relative initial positions determine the places of conjunctions of the two planets. If they meet regularly near the pericentre of the giant planet, the motion of the test planet becomes chaotic, while it can remain regular if the conjunctions take place near the apocenter of the giant planet. The effect of the initial phase difference between the planets is important, therefore we prepared stability maps for more initial values of the mean anomaly $M_1$ of the giant planet. The essential role of the phases for computations of orbits in mean motion resonances for terrestrial planets in exoplanetary systems has already been stressed by \cite{asg04}, where the giant planet was initially in periastron or apastron, and the phase of the terrestrial planet has been set to 12 different initial angles of the mean anomaly. \citep[A detailed treatment of the geometry of the resonances can be found in][]{mur99}.

By studying the series of Figs. \ref{fig6} - \ref{fig10}, one can follow the effect of the increase of the mass parameter on the stability structure. For larger mass of the giant planet, thus for larger perturbations, the chaotic region is more extended, as expected.

We note also a peculiar feature of  Figs. \ref{fig6}a, \ref{fig7} -- \ref{fig10}. In the strongly chaotic regions on the right, which are due to the proximity of the giant planet, there are two almost parallel light stripes, corresponding to satellite-type orbits around the giant planet. These orbits are less chaotic (or they may be ordered), than those emanating from the surrounding region. Between these stripes the dark gap corresponds to the perihelion distance of the giant planet, therefore in orbits originating from here, the test particles suffer close encounters with the giant planet, or they even can collide with it.

\section{Application of the catalogue to individual exoplanetary systems}
In this section we show how to use the catalogue to determine the stability of hypothetical Earth-like planets in exoplanetary sytems. As an example we consider the case of HD10697, where $a_1=2.13$ AU, $e_1=0.11$, $\mu=0.0055$. Fig. \ref{fig9} shows a stability map, calculated for $\mu=0.005$ for inner orbits of the test planet. This corresponds to the minimum mass of the giant planet (minimum masses are used throughout). The stability of a small planet (starting with $e=0$) in the system HD10697 can be studied along the line $e_1=0.11$. One can see, that for small semi-major axes, $a<0.33 a_1 (=0.729 \mbox{AU})$  the parameter space is very stable. When $a>0.33 a_1$ several resonances appear, among which the most important are the $5:1$, $4:1$, $3:1$, and $2:1$ resonances. For $a>0.73 a_1 (=1.55\mbox{AU})$  a strongly chaotic region appears. The classical HZ of this system is between $0.85-1.65$ AU \citep{men03}. Thus in Fig. \ref{fig9} the scaled classical HZ is situated between $0.85/a_1=0.39\ \mbox{and}\ 1.65/a_1=0.77$ (shown as a rectangle, elongated in horizontal direction). One can see, that the inner part of the classical HZ contains ordered regions, but stripes of certain resonances are also present. The outer part of the classical HZ is in the strongly chaotic region.

In Table \ref{tbl-1} we list the orbital parameters of the exoplanetary systems displayed in Figs. \ref{fig1}--\ref{fig10}, and we characterize the stability properties of their hypothetical terrestrial planets in their classical and present HZs given by \cite{men03} and \cite{jon06}. We give also the boundaries of the stable and the strongly chaotic regions. As sources for the parameters  we used the webpages \texttt{http://www.exoplanets.org} and Jean Schneider's catalogue (\texttt{http://exoplanet.eu}), denoted by E and JS, respectively, in Table 1 and in the figures, when there were significant differences between the data. In the column MMRs we display the highest and the lowest order mean motion resonances of the given exoplanetary system (not only in their HZ), the other resonances between them can be taken from the corresponding figure. When the highest order resonance is above $9:1$, we call it as h.o.r. In the column CHZ/PHZ we give a characterization of the stability properties of the classical and present HZs. The notation S means a fully stable HZ, where no mean motion resonance exists. When the HZ contains a few resonances, we consider it partly stable (PS). If many resonances fill in the HZ, we call it marginally stable (MS). Strongly chaotic, thus very unstable domains are denoted by SC. 

The uncertainty in the eccentricity $e_1$ of a giant planet can affect the stability of the HZ significantly. The uncertainty in $e_1$  is indicated on the stability maps by the width of the rectangle of the classical HZ. (In the case of the present HZ, marked by parentheses, we did not show the uncertainty in $e_1$, since it is the same.) When the uncertainty in $e_1$ is large, the stability properties at the top and at the bottom of the rectangle of the classical HZ can be quite different. For instance, in the case of the system HD 30177 it can be seen from Fig. 10, that for small values of $e_1$ there are a few resonances in the classical HZ, thus it is partly stable, while for high values of $e_1$ the HZ is full of resonances, therefore it is only  marginally stable.

\section{Comparison with other results}

The habitability of individual exoplanetary systems has been studied by many authors. As we have already mentioned, the orbital parameters of these systems are changed quite frequently, so it is not an easy task to compare directly our results with the results of other authors using older or different orbital data. On the other hand, there are only a few works studying the habitability of a large portion of the known systems. Such global investigations have been made by \cite{men03} studying the habitability of 85 known exoplanetary sytems, and more recently by \cite{jon06} investigating the present habitability of 152 exoplanetary sytems.

In these studies the gravitational influence of the known giant planet on the HZ is estimated mainly by some multiple of its Hill radius. However, the stability properties of the HZ of an exoplanetary system are also affected by mean motion resonances. If the HZ of a system contains only a few resonances, the ordered regions between them can provide stability for an Earth-like planet. If the resonances cover densely the HZ, the long term stability of an Earth-like planet is more questionable; either the resonances overlap each other generating a strongly chaotic region, or chaotic diffusion, acting inside a mean motion resonance, can destabilize the motion of a hypothetical small planet. We note that the presence of additional, yet undetected giant planets in an extrasolar planetary system can enhance the depletion of the mean motion resonances through their possible secular interactions with the known giant planet, similarly to the formation of the Kirkwood gaps in our Solar system \citep{mor93,moo95}.

In our stability maps we investigated the dynamical stability of the classical HZ given by \cite{men03}, and the present HZ calculated by \cite{jon06}.
Studying the dynamical stability of the classical and the present HZs of 15 extrasolar planetary systems (displayed in Table 1), we have found quite good agreement with the results of \cite{men03} and \cite{jon06}. We emphasize, however, that our stability maps also exhibit the resonant structure of the HZs, which is essential to establish the real stability properties of an Earth-like planet.

In what follows we characterize the HZ of some individual exoplanetary systems.
\begin{description}
\item[HD 52256 and HD 121504:]
Depending on the data source, the classical HZ of the system HD 121504 is either fully stable (source E), or partly stable (source JS), containing a few resonances (Fig. \ref{fig1}). The classical HZ of the system HD 52256 is partly stable, but contains more resonances and smaller stability regions for the data JS, than for E. The present HZs of these systems are located in the very regular regions of the $(a,e_1)$ parameter plane with a few resonances in the case of HD 52256 for the data JS. Our results are in good agreement with the results of \cite{men03} and \cite{jon06} concerning the extension of the stability regions. These systems could harbor stable habitable Earth-like planets.
\begin{figure}
\includegraphics[width=0.95\linewidth]{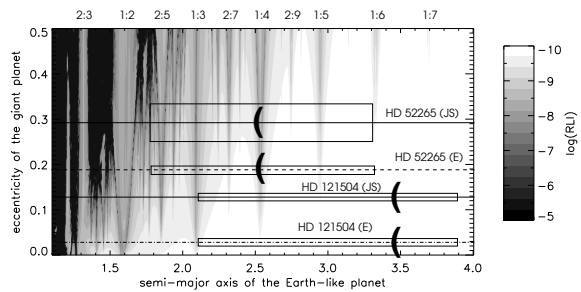}
\caption{Stability map, computed by the RLI for outer orbits of the test planet, when $\mu=0.001$ and $M_1=0^{\circ}$. The classical HZs of two exoplanetary sytems (with orbital parameters taken from two different data sources, denoted by E and JS, see Section 5) are also indicated as elongated rectangles, whose width corresponds to the uncertainty in the eccentricity of the giant planet's orbit of each system. The inner borders of the present HZs are displayed as opened parentheses. The outer borders are out of the investigated range of the $(a,e_1)$ plane. As described in Sections 5 and 6, it can be seen, that depending on the data source the classical HZ of the system HD 121504 is either fully stable (source E), or partly stable (source JS), containing a few resonances. The system HD 52265 is partly stable, but contains more resonances and smaller regions of stability for the data JS, than for E. The present HZs are located in the very regular regions of the $(a,e_1)$ plane.}
\label{fig1}
\end{figure}
\begin{figure}
\includegraphics[width=0.95\linewidth]{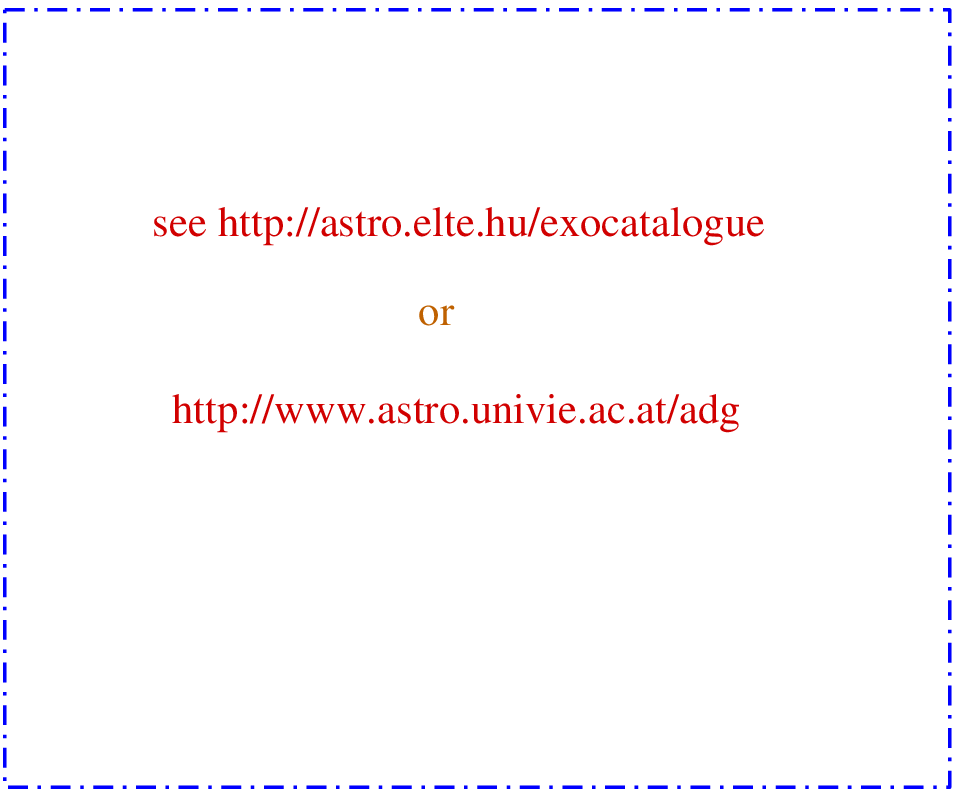}
\caption{Stability map, computed by the FLI for outer orbits of the test planet, when $\mu=0.001$ and $M_1=0^{\circ}$. Light shades indicate stable regions, dark means chaoticity. The very light stripes correspond to stable resonant orbits in the mean motion resonances. Note the complementary character of Figs. \ref{fig1} and \ref{fig2}, described in the text.}
\label{fig2}
\end{figure}
\begin{figure}
\includegraphics[width=0.95\linewidth]{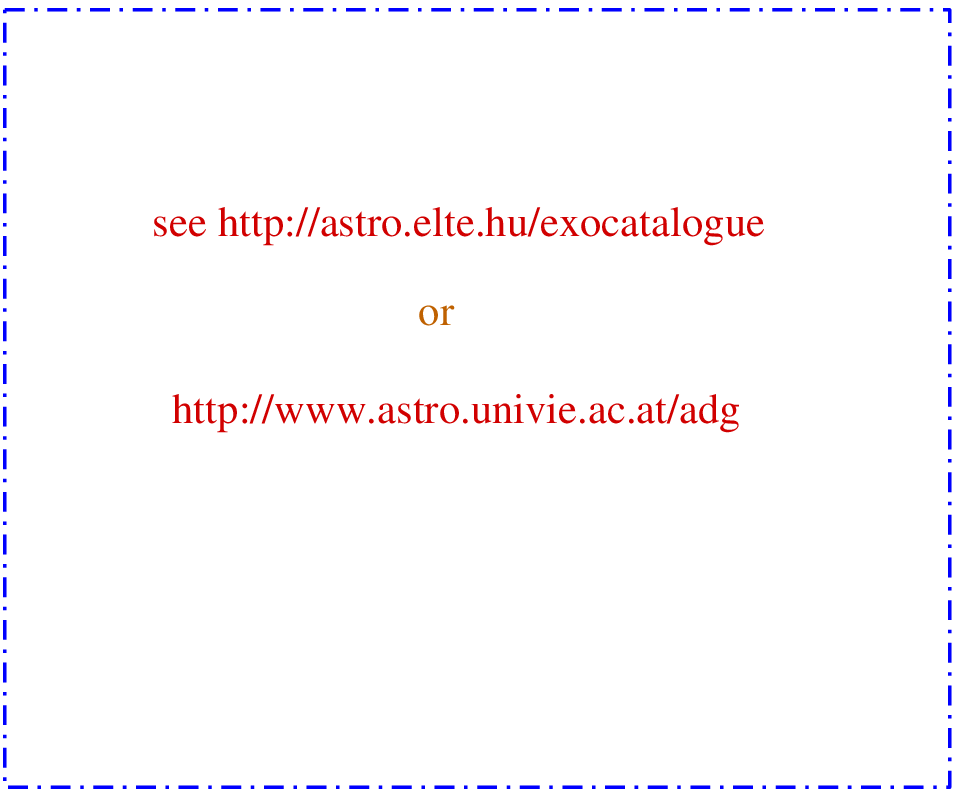}
\caption{Stability map, computed by the MEM for outer orbits of the test planet, when $\mu=0.001$ and $M_1=0^{\circ}$. Here again lighter shades mean orbits, where the eccentricity remains relatively small. The changes in the maximum eccentricity of the test planet indicate clearly the positions of the mean motion resonances.}
\label{fig3}
\end{figure}

\item[HD 8574:]
The classical HZ of this system is in the strongly chaotic region for both data sources E and JS (Fig. \ref{fig4}). \cite{jon06} found that $72\%$ of the present HZ of the system can survive the gravitational influence of the giant planet, thus it could contain a stable habitable planet. Our approach clearly shows that the present HZ of the system is full with overlaping mean motion resonances, and only partly stable for the data E, and marginally stable for the data JS. Thus the chance for the system to host a stable habitable planet is small. 
\begin{figure}
\includegraphics[width=0.95\linewidth]{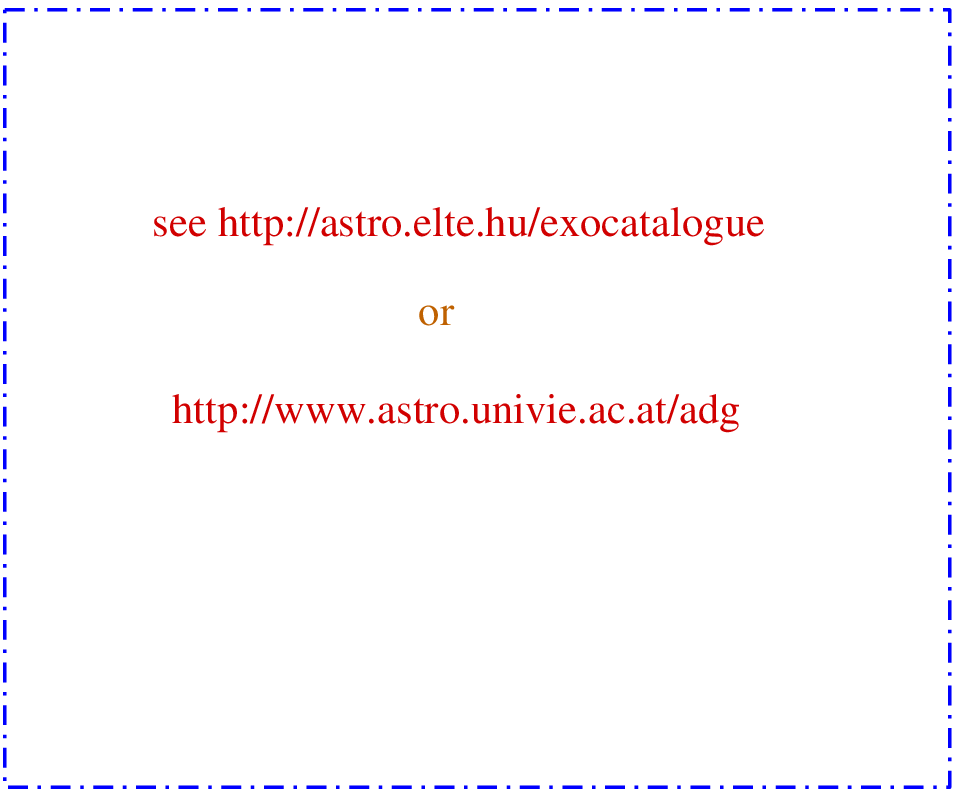}
\caption{Stability map, computed by the RLI for outer orbits of the test planet, when $\mu=0.002$ and $M_1=0^{\circ}$. A comparison with Fig. 5 reveals the increase of the chaotic region for increasing $\mu$. As described in Section 6, the classical HZ of the system HD 8574 is strongly chaotic for both data sources E and JS. The present HZ of the system (denoted by parentheses) is partly stable for E, and marginally stable for JS.}
\label{fig4}
\end{figure}

\item[70 Virginis and HD 178911 B:]
The classical HZ of 70 Vir is partly in the strongly chaotic region
(Fig. \ref{fig5}), while its remaining part is full with overlaping mean motion resonances and thus only marginally stable. This means that the classical HZ of this system is unlikely to host habitable Earth-like planets. The classical HZ of HD 178911 B is partly stable, and it can contain stable Earth-like planets. Our results are in good agreement with the results of \cite{men03}. The present HZs of these systems are in the ordered regions of the $(a,e_1)$ plane, thus these systems can host Earth-like habitable planets in stable orbits. These results agree well with the results of \cite{jon06}.
\begin{figure}
\includegraphics[width=0.95\linewidth]{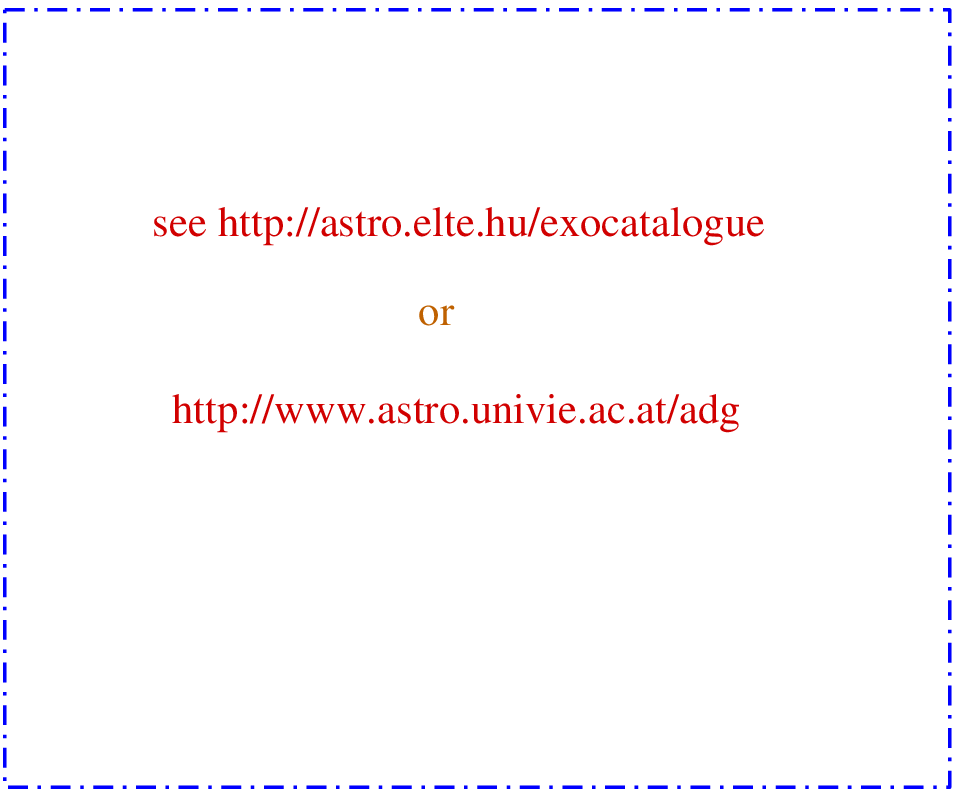}
\caption{Stability map, computed by the RLI for outer orbits of the test planet, when $\mu=0.007$ and $M_1=0^{\circ}$. As described in Section 6, the classical HZ of 70 Vir is partly in the strongly chaotic region, while its other part is marginally stable. The classical HZ of the system HD 178911 B is partly stable. The present HZ of  HD 178911 B (denoted by parenthesis) is stable, while at 70 Vir it is partly stable, containing higher order mean motion resonances (1:8, 1:9, etc., not shown in the figure).}
\label{fig5}
\end{figure}

\item[Eps Eri, HD 114729, and GJ 777 A:]
The classical HZs of Eps Eri and HD 114729 are marginally stable (Fig. \ref{fig6}). There is only a small chance to find Earth-like planets here. \cite{jon06} found that only 17\% of the present HZ of Eps Eri can have stable orbits, and nowhere in HD114729 does so. In the case of GJ 777 A, due to the large uncertainty in $e_1$ of the giant planet ($e_1=0.48\pm0.2$), the classical HZ can be stable (for $e_1=0.48-0.2$), or partly stable (for $e_1=0.48$). There are no computed stability data for $e_1>0.5$, however, the stable region will certainly decrease for larger values of $e_1$. The present HZ of GJ 777 A for $e_1=0.48$ is partly in the strongly chaotic region of the $(a,e_1)$ plane, while its remaining part is filled by overlaping mean motion resonances and is marginally stable. For $e_1=0.28$, the present HZ of GJ 777 A is partly stable. \cite{jon06} found that 64\% of the present HZ can survive the gravitational influence of the giant planet. However, even though resonances were included in the development of their fast method of assessing stability \citep{jon05}, the method does not permit detailed consideration of the resonances in the individual systems to which it was applied. Considering the classical HZ, this system can host stable Earth-like planets, as has also been found by \cite{men03} and \cite{asg04}.
\begin{figure}
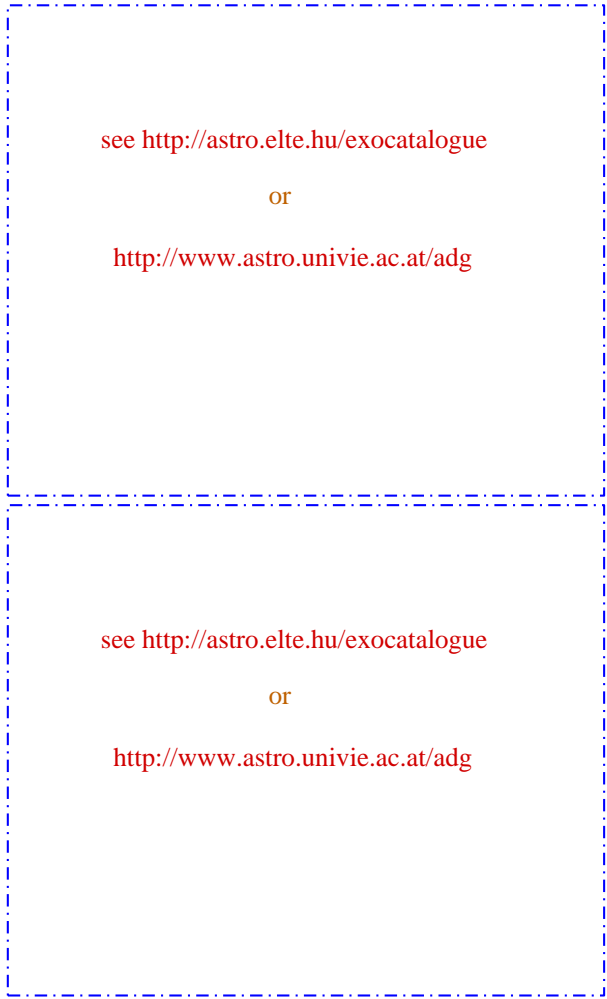

\includegraphics[width=0.95\linewidth]{x.eps}
\includegraphics[width=0.95\linewidth]{x.eps}
\caption{Stability maps, computed by the RLI for inner orbits of the test planet. (a) $\mu=0.001$, $M_1=0^{\circ}$; (b) $\mu=0.001$, $M_1=180^{\circ}$. The comparison of the two figures shows the effect of the relative starting positions of the two planets (for more details see the text in Section 4). As described in Section 6, the classical HZs of Eps Eri and  HD 114729 are marginally stable. Due to the large uncertainty of the giant planet's eccentricity, the classical HZ of GJ 777 A can be stable (for $e_1=0.28$), or partly stable (for $e_1=0.48$). The HZ of the Solar System is placed on the map under the approximate condition that there is only one giant planet, Jupiter in the system. It is reassuring that the HZ of the Solar System is in the stable region. The present HZ of GJ 777 A for $e_1=0.48$ is partly in the chaotic region, while its other part is marginally stable, and for $e_1=0.28$ it is partly stable.}
\label{fig6}
\end{figure}

\item[HD 70642:]
The classical HZ of this system can be stable, or partly stable, depending on the eccentricity of the giant planet (Fig. \ref{fig7}). The present HZ of the system contains a few mean motion resonances, but between them there are ordered regions, which can host stable Earth-like planets. This result has also been found by \cite{jon06}. 
\begin{figure}
\includegraphics[width=0.95\linewidth]{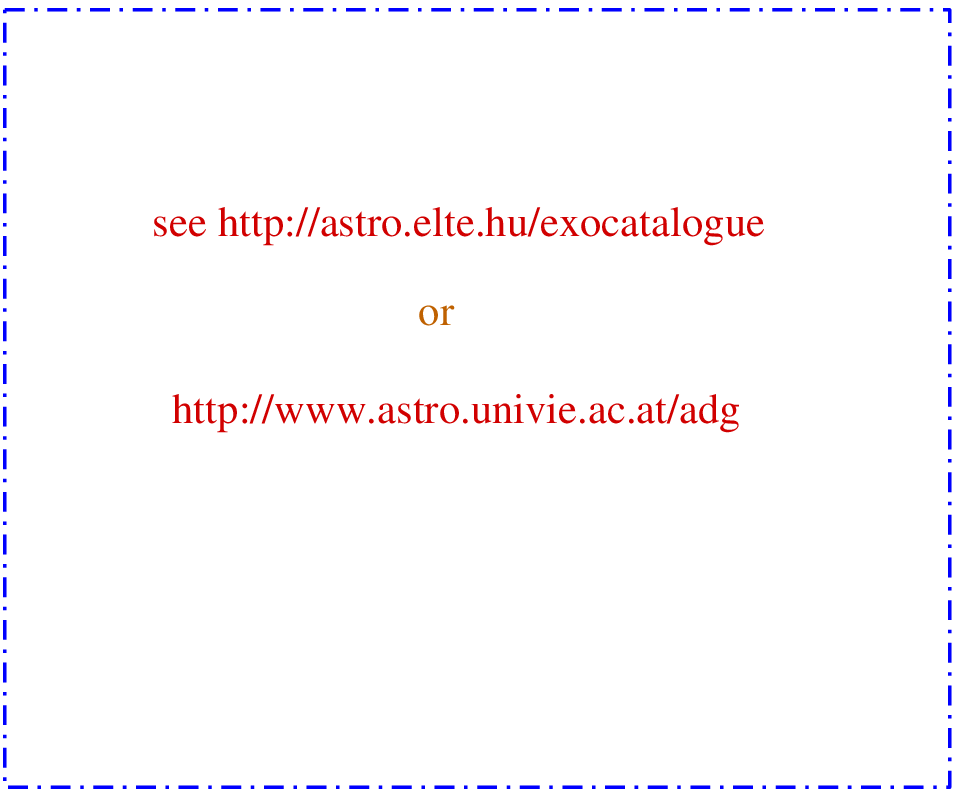}
\caption{Stability map, computed by the RLI  for inner orbits of the test planet, when $\mu=0.002$ and $M_1=0^{\circ}$. Due to the large uncertainty of the eccentricity of the giant planet, the classical HZ of the system HD 70642 is either stable, or partly stable. The present HZ of this system is partly stable within the lower and upper limits of $e_1$.}
\label{fig7}
\end{figure}

\item[HD 72659:]
The classical HZ of this system contains only a few mean motion resonances (Fig. \ref{fig8}), thus it can host stable Earth-like planets. This is in good agreement with the results of \cite{men03} and \cite{asg04}. \cite{jon06} found, that $71\%$ of the present HZ is not disrupted gravitationally by the giant planet. This is true, if one considers only the effect of the strongly chaotic region. However, the inner part of the present HZ is full with mean motion resonances and is marginally stable. Thus the place for a stable Earth-like planet is much smaller, than $71\%$ of the whole present HZ, when the effects of the mean motion resonances are taken into consideration. However, the presence of an Earth-like planet cannot be excluded entirely.
\begin{figure}
\includegraphics[width=0.95\linewidth]{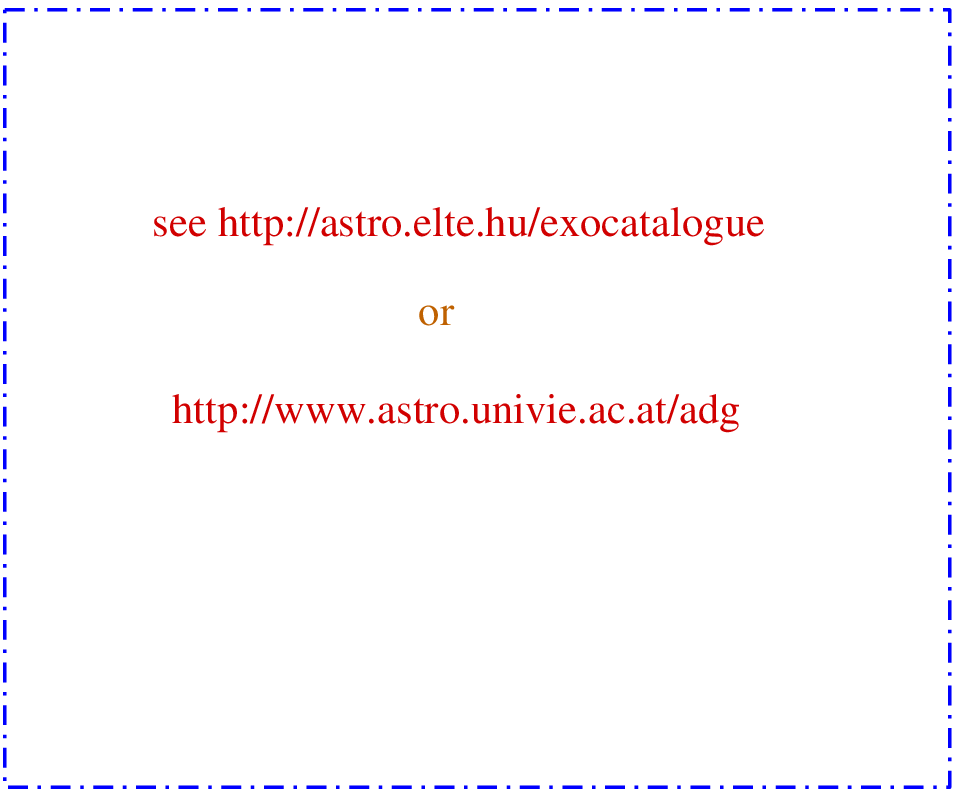}
\caption{Stability map, computed by the RLI  for inner orbits of the test planet, when $\mu=0.003$ and $M_1=0^{\circ}$. The classical HZ of the system HD 72659 is partly stable. The outer part of the present HZ of the system is chaotic, while its inner part is marginally stable. We placed also the classical HZ of the multiple planet system 47 UMa on the figure for comparison, by taking into account from its two giant planets the more massive component b.}
\label{fig8}
\end{figure}

\item[HD 50554:]
Both the classical and the present HZ of this system are entirely in the strongly chaotic region of the $(a,e_1)$ plane (Fig. \ref{fig9}). Thus this system can not host habitable Earth-like planet. \cite{jon06} came to the same conclusion. On the other hand, there could be (retrograde) satellite-type orbits in the system.  

\item[14 Her:]
Considering the classical HZ, this system can host stable Earth-like planets, however, mean motion resonances fill densely the classical HZ (Fig. \ref{fig9}). As to the present HZ, the place for a stable Earth-like planet is much smaller, with one part in the chaotic region, and the other is filled by many resonances. Our results are in good agreement with the results of \cite{jon06}.

\item[HD 10697:]
The classical HZ of this system contains a strongly chaotic region, while its inner part between the mean motion resonances contains ordered, thus stable regions as well (Fig. \ref{fig9}). The present HZ of this system is in the strongly chaotic region, thus it cannot harbour stable Earth-like planets. This is also found by \cite{jon06}.

\begin{figure}
\includegraphics[width=0.95\linewidth]{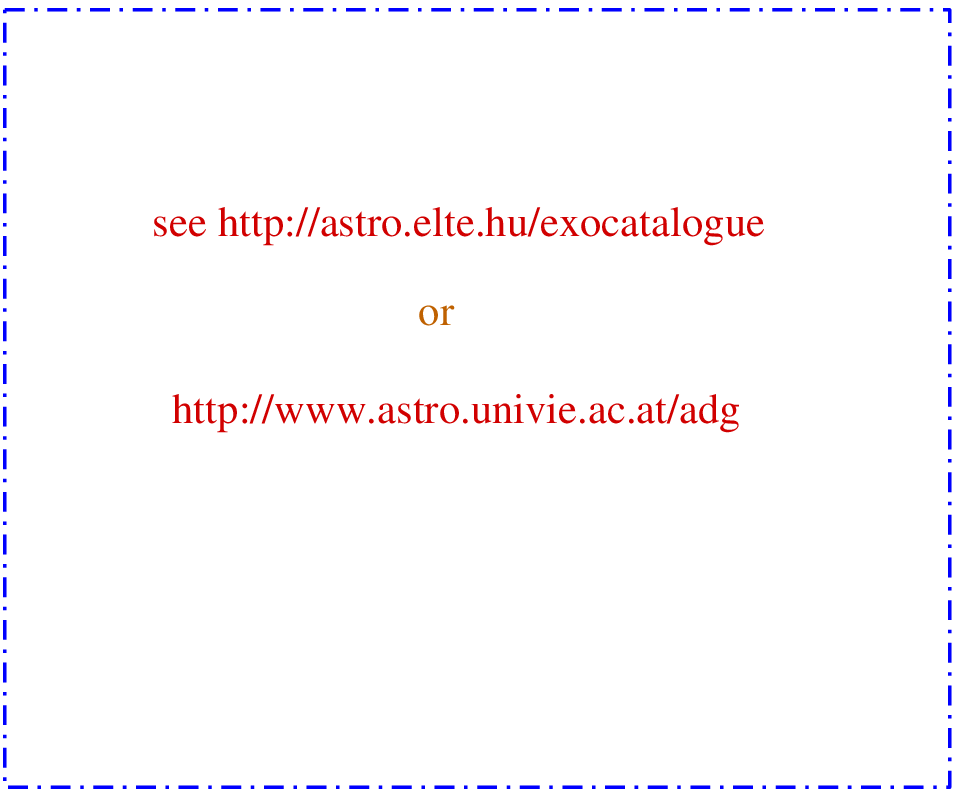}
\caption{Stability map, computed by the RLI  for inner orbits of the test planet, when $\mu=0.005$ and $M_1=0^{\circ}$. Along the line $e_1=0.11$, corresponding to the system HD 10697, the parameter space is very stable for $a<0.33$. For $a>0.73$ the parameter space is strongly chaotic. Between $0.33<a<0.73$ there are several resonances. The scaled classical HZ of the system HD 10697 is between $0.39<a<0.77$, thus its inner part is partly stable, but its outer edge is strongly chaotic. So is its present HZ. As described in Section 6, both the classical and the present HZ of the system HD 50554 are strongly chaotic. Interestingly, there could be (retrograde) satellite-type orbits in this system. The classical HZ of 14 Her is partly stable, while its present HZ is chaotic on one side, and marginally stable on the other.}
\label{fig9}
\end{figure}

\item[HD 30177:]
The classical HZ of this system may host Earth-like planets, since between the mean motion resonances there are ordered regions on the $(a,e_1)$ parameter plane (Fig. \ref{fig10}). Considering the present HZ, its outer part is in the strongly chaotic region, while the inner part of it may contain stable Earth-like planets. This result agrees well with the result of \cite{jon06}.
\begin{figure}
\includegraphics[width=0.95\linewidth]{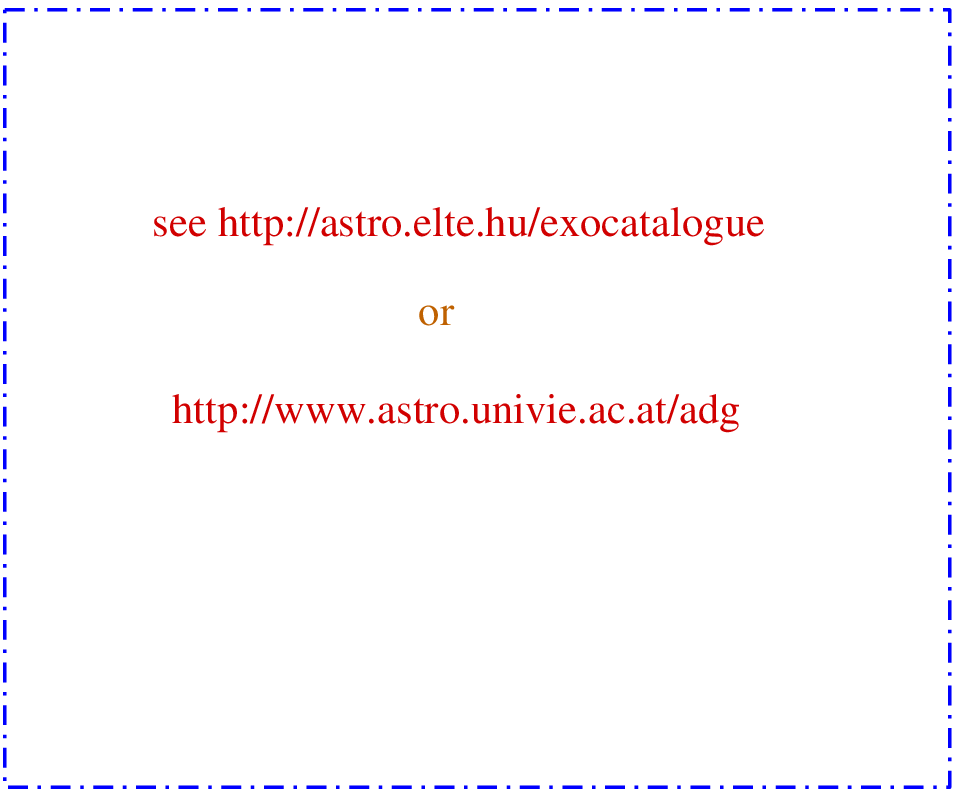}
\caption{Stability map, computed by the RLI for inner orbits of the test planet, when $\mu=0.009$ and $M_1=0^{\circ}$. Due to the large uncertainty in the eccentricity of the giant planet, the classical HZ of the system HD 30177 is either partly stable (at the bottom of the rectangle), or marginally stable (at the top of the rectangle). The present HZ is strongly chaotic for large values of $e_1$, while its inner part is partly or marginally stable. }
\label{fig10}
\end{figure}
\end{description}

Thus we can conclude, that the effect of the mean motion resonances cannot be neglected, when studying the habitability of the known extrasolar planetary systems. In this sense our work complements the already existing studies, such as \cite{jon06}.

\section{Summary}

In this paper we described a catalogue of stability maps of hypothetical terrestrial planets in exoplanetary systems with one giant planet. By using these maps, one can easily study the stability properties of possible Earth-like planets and the habitable zones in known or future exoplanetary systems. The novelty of our work is the invariance of our results with respect to the frequent changes in the computed orbital parameters of the extrasolar planetary systems. Thus any new orbital determination (leading to new elements) can immediately be checked without any time-consuming computations. This catalogue could also be useful for observing astronomers in determining the orbital parameters of terrestrial planets detected in the future by space instruments (COROT, KEPLER) using transit photometry.

\begin{table*}
\begin{minipage}{126mm}
\caption{Parameters of the extrasolar planetary systems displayed in Figs. \ref{fig1}--\ref{fig10}. The semi-major axis $a_1$ and the boundaries of the stable and the strongly chaotic regions are in AU. MMR means mean motion resonance, CHZ stands for the classical, PHZ for the present habitable zone. The characterization is: S stable, PS partly stable, MS marginally stable, SC strongly chaotic (n.d. means, that there are no data for the present HZ). The data sources E and JS are explained in the text.}
\label{tbl-1}
\begin{tabular}{lcccccccl}
\hline\hline
System & $a_1 $ & $e_1$ &stable &MMRs &strongly chaotic & CHZ/ \\
& & & region & & region & PHZ \\
\hline
HD 52265 (E) & 0.49 & 0.19 &$a>1.47$& 2:3$\rightarrow$1:5& $a<0.735$&PS/S\\
\hline
HD 52265 (JS) & 0.49 & 0.29 &$a>1.47$& 2:3$\rightarrow$1:6& $a<0.735$&PS/PS\\
& & & & & & &  \\
\hline
HD 121504 (E) & 0.33 & 0.03 & $a>0.69$& 2:3$\rightarrow$1:3& $a<0.429$& S/S\\
\hline
HD 121504 (JS) & 0.33 & 0.13 & $a>0.82$& 2:3$\rightarrow$1:5& $a<0.49$& PS/S\\
\hline
HD 8574 (E) & 0.77 & 0.288 & $a>2.618$& 2:3$\rightarrow$1:6 & $a<1.54$& SC/PS\\
& &$\pm$ 0.05 & & & & &  \\
\hline
HD 8574 (JS) & 0.77 & 0.4 & $a>2.85$& 2:3$\rightarrow$1:7 & $a<1.54$& SC/MS\\
& &$\pm$ 0.04 & & & & &  \\
\hline
70 Vir & 0.48 & 0.4 & $a>1.82$& 2:3$\rightarrow$1:7 & $a<1.04$& MS/PS\\
\hline
HD 178911 B& 0.32 & 0.12 & $a>0.99$& 1:3$\rightarrow$1:6 & $a<0.68$& PS/S\\
\hline
Eps Eri & 3.4 &0.43  & $a < 0.58$ & h.o.r. $\rightarrow$ 4:1 & $a > 1.36$ & MS/n.d. \\
\hline
HD 114729 & 2.08&0.31 &$a<0.47$&8:1$\rightarrow$3:1&$a>1.04$& MS/n.d. \\
\hline
GJ 777 A & 4.8 &$0.48$&$a<0.82$ & h.o.r$\rightarrow$7:3&$a>1.63$& PS/MS\\
              &       &$0.48-0.2$ &$a<1.10$ &            $9:1, 8:1$        &$a>2.69$& S/PS\\
\hline
HD 70642& 3.3 & 0.1+0.06 & $a<0.89$& 6:1$\rightarrow$2:1 & $a>2.24$ & PS/PS\\
                 &       &0.1 - 0.06 & $a<1.32$&         5:1, 4:1            & $a>2.57 $ & S/PS \\
\hline
HD 72659& 4.16&     0.2     & $a<1.04$& 9:1$\rightarrow$7:3& $a>3.12$ & PS/MS\\
\hline
47UMa b  & 2.09 &    0.06   & $a<0.71$& 5:1$\rightarrow$2:1& $a>1.52$ & PS/\\
\hline
HD 50554&2.38 &    0.42   & $a<0.24$& h.o.r$\rightarrow$5:1& $a>0.83$ & SC/SC\\
                 &        &$\pm$0.03&      &                                 &          &      \\
\hline
14 Her &2.85&0.38&$a<0.59$ &h.o.r.$\rightarrow$4:1 & $a>0.99$&PS/MS \\
\hline
HD 10697 &2.13&0.11&$a<0.72$&5:1$\rightarrow$2:1&$a>1.4$ &PS/SC \\
\hline
HD 30177 &3.86&$0.3+0.17$&$a<0.19$&h.o.r.$\rightarrow$7:1&$a>1.16$&MS/SC\\
                  &       & $0.3-0.17$&$a<1.16$&h.o.r.$\rightarrow$7:3&$a>2.9$&PS/PS\\
\hline\hline
\end{tabular}
\end{minipage}
\end{table*}

\section*{Acknowledgments}

ZsS, \'AS, and B\'E thanks the support of the Hungarian Scientific Research Fund (OTKA) under the grants D048424 and T043739. EP-L wishes to acknowledge the support by the Austrian FWF (Hertha Firnberg Project T122). This research has also been supported by the Austrian-Hungarian Scientific and Technology Cooperation under the grant A-12/04, and also benefited fom the ISSI team "Evolution of Habitable Planets". The numerical integrations were done on the NIIDP supercomputer in Hungary. We thank the referee for useful suggestions.

\end{document}